\documentclass[lettersize,journal]{IEEEtran}
\usepackage{amsmath,amsfonts}
\usepackage{algorithmic}
\usepackage{algorithm}
\usepackage{array}
\usepackage[caption=false,font=normalsize,labelfont=sf,textfont=sf]{subfig}
\usepackage{textcomp}
\usepackage{stfloats}
\usepackage{url}
\usepackage{verbatim}
\usepackage{graphicx}
\usepackage{cite}
\usepackage{booktabs}
\usepackage{multirow}
\usepackage{fontawesome}
\usepackage{cleveref}
\usepackage{tabularx}
\usepackage{comment}
\usepackage{makecell}
\usepackage{adjustbox}
\usepackage[table]{xcolor}
\usepackage{colortbl}
\usepackage{color}
\usepackage{multicol}
\usepackage{multirow}
\crefname{section}{§}{§§}
\Crefname{section}{§}{§§}
\usepackage{tabularray}
\usepackage{multirow}
\begin{document}

\title{Generative AI for Vulnerability Detection in 6G Wireless Networks: Advances, Case Study, and Future Directions}

\author{Shuo Yang, Xinran Zheng, Jinfeng~Xu, Jinze Li, Danyang Song, Zheyu Chen and Edith~C.H.~Ngai$^{*}$

\thanks{S. Yang, J. Xu, J. Li, D. Song, and E. C.H. Ngai are with Department of Electrical and Electronic Engineering, The University of Hong Kong, Hong Kong ASR, China (shuo.yang@connect.hku.hk).}
\thanks{X. Zheng is with Department of Electronic Engineering, Tsinghua University, Beijing, China.}
\thanks{Z. Chen is with Department of Electrical and Electronic Engineering, The Hong Kong Polytechnic University, Hong Kong ASR, China}
\thanks{$^{*}$ Corresponding author: Edith~C.H.~Ngai (chngai@eee.hku.hk).}
}

\markboth{Journal of \LaTeX\ Class Files,~Vol.~14, No.~8, August~2021}%
{Shell \MakeLowercase{\textit{et al.}}: A Sample Article Using IEEEtran.cls for IEEE Journals}

\maketitle

\begin{abstract}

The rapid advancement of 6G wireless networks, IoT, and edge computing has significantly expanded the cyberattack surface, necessitating more intelligent and adaptive vulnerability detection mechanisms. Traditional security methods, while foundational, struggle with zero-day exploits, adversarial threats, and context-dependent vulnerabilities in highly dynamic network environments. Generative AI (GAI) emerges as a transformative solution, leveraging synthetic data generation, multimodal reasoning, and adaptive learning to enhance security frameworks. This paper explores the integration of GAI-powered vulnerability detection in 6G wireless networks, focusing on code auditing, protocol security, cloud-edge defenses, and hardware protection. We introduce a three-layer framework—comprising the Technology Layer, Capability Layer, and Application Layer—to systematically analyze the role of VAEs, GANs, LLMs, and GDMs in securing next-generation wireless ecosystems. To demonstrate practical implementation, we present a case study on LLM-driven code vulnerability detection, highlighting its effectiveness, performance, and challenges. Finally, we outline future research directions, including lightweight models, high-authenticity data generation, external knowledge integration, and privacy-preserving technologies. By synthesizing current advancements and open challenges, this work provides a roadmap for researchers and practitioners to harness GAI for building resilient and adaptive security solutions in 6G networks.

\end{abstract}


\section{Introduction}
\label{sec: intro}

\begin{figure}
    \centering
    \includegraphics[width=1\linewidth]{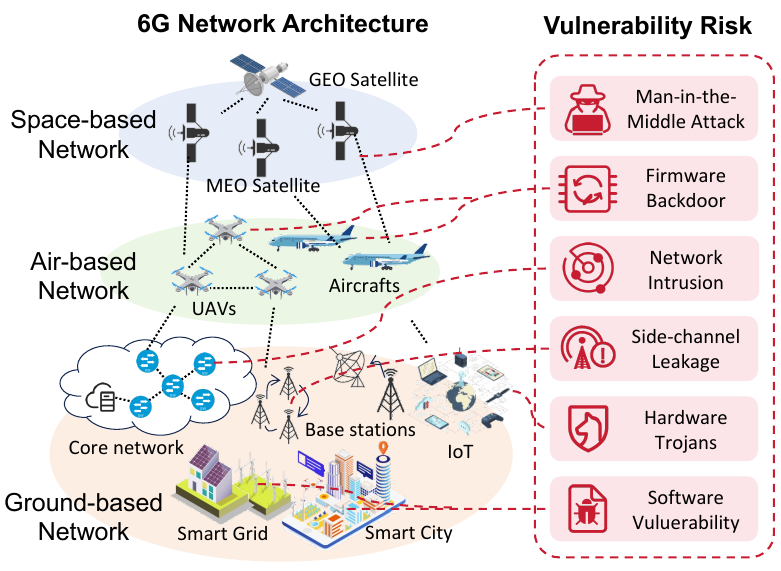}
    \caption{The architecture of 6G wireless networks and some of the associated vulnerability risks}
    \label{fig: 6g}
\end{figure}

Wireless networks form the backbone of modern connectivity, enabling innovations in smart cities, autonomous systems, and Industry 4.0. However, their growing complexity—characterized by heterogeneous architectures, dynamic topologies, and resource-constrained edge devices—has introduced unprecedented security challenges~\cite{survey}. The increasing reliance on AI-driven network orchestration, ultra-low latency communications, and decentralized computing in 6G environments further exacerbates security risks, making traditional vulnerability detection methods inadequate for emerging threats. As illustrated in Fig.~\ref{fig: 6g}, the multi-layered architecture of the 6G wireless network incorporates space-based, air-based, and ground-based networks. While this architecture enables ubiquitous connectivity and advanced services, it also exposes the network to a variety of vulnerability risks. For example, attackers may intercept or manipulate data transmitted through terrestrial or inter-satellite links, while hardware Trojans could be implanted during manufacturing or maintenance in chips and communication modules, enabling remote control or data theft. Additionally, the operating systems and communication protocol stacks of UAVs, ground stations, and other devices may harbor unpatched firmware backdoors. The heterogeneity and vast scale of these interconnected domains significantly complicate the task of comprehensive vulnerability detection and mitigation.

Traditional vulnerability detection methods, which rely on static signatures, expert-crafted rules, and manual analysis, struggle to detect zero-day exploits, adversarial threats, and context-dependent vulnerabilities~\cite{grace, agent, cloud}. These limitations are compounded by the vast scale of wireless ecosystems, proprietary firmware complexities, and opaque protocol implementations. As attackers increasingly employ AI-generated adversarial attacks and malware, traditional detection systems are unable to adapt or scale effectively to the rapidly evolving threat landscape.

Generative AI (GAI) offers a paradigm shift in vulnerability detection by integrating synthetic data generation~\cite{trojanforge}, semantic reasoning~\cite{cloud}, and adaptive security modeling~\cite{agent}. Techniques such as Generative Adversarial Networks (GANs) facilitate attack simulation and adversarial stress testing~\cite{trojanforge}, while Large Language Models (LLMs) enable deep semantic analysis of code and protocol specifications~\cite{llmif}. By learning from diverse multimodal datasets—including network traffic, hardware logs, and behavioral records—GAI enhances coverage, real-time adaptability, and human-AI collaboration, enabling more effective cybersecurity frameworks for 6G-enabled wireless infrastructures.

This paper provides a comprehensive exploration of GAI's transformative potential in enhancing wireless network security, introducing a three-layer framework that categorizes GAI's impact into technology, capability, and application layers, accompanied by a review of representative implementations. The technology layer focuses on foundational generative models, including Variational Autoencoders (VAEs), GANs, LLMs, and Generative Diffusion Models (GDMs). The capability layer explores core GAI-driven functionalities, which encompass content generation, multimodal modeling, semantic reasoning, adaptive learning, and human-AI collaboration. The application layer delves into real-world implementations of GAI across various domains in wireless networks, including code security, protocol analysis, cloud-edge defense, and hardware protection.

To illustrate practical implementation, this paper presents a case study focusing on code security, specifically conducting LLM-driven code vulnerability detection, analyzing its effectiveness, performance metrics, and inherent limitations. Additionally, we outline future research directions aimed at further leveraging GAI to enhance wireless security, including lightweight models, high-authenticity data generation, adversarial robustness, external knowledge integration, and privacy-preserving technologies. In summary, this survey paper seeks to inspire innovative applications of Generative AI (GAI) within the 6G landscape. It highlights GAI's transformative potential, open issues, and provides a pathway toward achieving adaptive, explainable, and reliable cybersecurity in 6G ecosystems.

The remainder of this paper is structured as follows: Section~\ref{sec: background} provides an overview of traditional vulnerability detection pipelines and their limitations in wireless networks. Section~\ref{sec: method} introduces the proposed GAI-enabled vulnerability detection framework, detailing its technological foundations, core capabilities, and security applications. Section~\ref{sec: case} presents a case study on LLM-driven code vulnerability detection, illustrating the practical impact of GAI in cybersecurity. Section~\ref{sec: future} outlines future research directions, and Section~\ref{sec: conclusion} concludes the paper.
\section{Background of Vulnerability Detection}
\label{sec: background}

Vulnerability detection is a fundamental component of cybersecurity, tasked with systematically identifying weaknesses in software, networks, and hardware to mitigate potential exploitations~\cite{vd}. This section outlines the traditional vulnerability detection pipeline and critically examines its limitations, particularly in addressing emerging security threats in next-generation wireless networks.

\subsection{Vulnerability Detection Pipeline}

\begin{figure}
    \centering
    \includegraphics[width=1\linewidth]{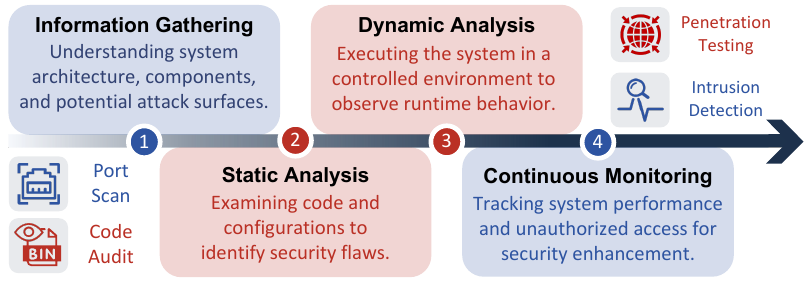}
    \caption{Vulnerability detection pipeline}
    \label{fig: pipeline}
\end{figure}

As illustrated in Fig.~\ref{fig: pipeline}, the general vulnerability detection process encompasses four phases: information gathering, static analysis, dynamic analysis, and continuous monitoring. These phases form the foundation for identifying, analyzing, and mitigating security vulnerabilities.

\subsubsection{Information Gathering}

The first phase involves comprehensive reconnaissance of system architecture, software components, dependencies, and network configurations to identify potential attack surfaces. Analysts utilize automated scanning tools alongside manual assessment techniques to extract metadata, detect open ports, and analyze exposed services. Tools such as Nmap\footnote{Nmap: https://nmap.org/} and Masscan\footnote{Masscan: http://github.com/robertdavidgraham/masscan} facilitate network reconnaissance by identifying accessible endpoints, running services, and protocol versions. Additionally, fingerprinting techniques determine software versions, helping correlate known vulnerabilities with deployed systems\footnote{Fingerprinting Tool: https://www.shodan.io/}. Effective information gathering ensures targeted vulnerability assessments, guiding subsequent phases toward efficient and precise detection~\cite{vd}.

\subsubsection{Static Analysis}

Static analysis involves inspecting source code, binaries, and configuration files without execution, allowing for the early identification of insecure coding patterns, misconfigurations, and logic flaws~\cite{grace}. This phase uses formal verification techniques to uncover potential security risks before deployment. Methods like taint analysis track untrusted data flow, abstract syntax tree (AST) inspection reveals structural weaknesses, and symbolic execution explores program paths to identify security flaws. Static Application Security Testing (SAST) tools automate vulnerability detection, enhancing efficiency in secure software development. However, static analysis can generate false positives~\cite{vd}, flagging issues that may not be exploitable in real-world scenarios, requiring further validation through dynamic analysis.

\subsubsection{Dynamic Analysis}

Dynamic analysis evaluates system behavior during execution, detecting vulnerabilities that manifest only at runtime, such as memory corruption, race conditions, and privilege escalation~\cite{vd}. This phase is essential for capturing real-time system interactions and identifying context-dependent attack vectors. Techniques such as fuzz testing generate malformed inputs to trigger unexpected behaviors, while penetration testing simulates real-world attack scenarios to uncover exploitable weaknesses. Sandboxing further enhances security assessments by isolating applications and monitoring their runtime activities for suspicious patterns. Dynamic analysis is particularly crucial in wireless security, where distributed systems, real-time communications, and large-scale infrastructures introduce high variability and unpredictability~\cite{survey}. However, the resource-intensive nature of dynamic analysis poses challenges in scalability and deployment feasibility within 6G network environments.

\subsubsection{Continuous Monitoring}

Given the evolving nature of cyber threats, continuous monitoring is essential for real-time detection and response. This phase integrates automated security mechanisms such as intrusion detection systems (IDS), network performance monitoring, and log anomaly detection to track system behavior and detect unauthorized activities. By analyzing network metrics, security frameworks can identify deviations from expected traffic patterns, enabling early threat mitigation. In wireless environments, where threats emerge dynamically across interconnected devices, continuous monitoring ensures proactive defense against evolving attack vectors. However, the effectiveness of continuous monitoring depends on its ability to minimize false positives~\cite{edge}, provide explainable alerts, and scale efficiently to complex network infrastructures.

\subsection{Limitations of Traditional Method with Wireless Threats}

Despite their widespread adoption, traditional vulnerability detection methods face significant limitations, particularly in wireless security of 6G networks. The increasing complexity of AI-driven security, ultra-dense connectivity, and intelligent edge computing introduces new challenges that conventional techniques struggle to manage effectively.

\subsubsection{Limited Coverage for Novel Threats}

Signature-based detection systems depend on predefined rules and heuristics, rendering them ineffective against zero-day exploits and new attack patterns~\cite{agent}. The rise of 6G wireless networks creates attack surfaces that conventional tools cannot address. For example, adversarial AI can evade traditional detection, while vulnerabilities in quantum-resistant encryption and intelligent edge computing may go undetected due to static and dynamic analysis limitations. As cyber threats surpass predefined detection capabilities, traditional methods struggle to ensure comprehensive security coverage.

\subsubsection{Inadequate Context Awareness}

Traditional methods analyze code, binaries, or runtime behavior in isolation, lacking the ability to correlate findings with system-wide security contexts. However, 6G environments, characterized by high-speed communications and the Internet of Everything, require context-aware security models. Static analysis may identify vulnerabilities based solely on code structure, ignoring real-world execution conditions, while dynamic analysis can detect anomalies but fails to correlate with broader network telemetry and behavioral intelligence. Without deeper contextual awareness, security mechanisms cannot accurately assess vulnerabilities in interconnected systems, resulting in inaccurate risk evaluations and increased false positives~\cite{grace}.

\subsubsection{Lack of Interpretability}

Many security tools produce complex, technical reports that are hard for analysts to interpret and act upon. Traditional static and dynamic analysis tools often generate an overwhelming number of alerts, lacking clear explanations about exploitability, potential impact, or mitigation strategies. Additionally, machine learning-based models act as black-box classifiers, providing little human-readable insight into why certain functions or behaviors are considered vulnerable~\cite{vd}. This lack of interpretability undermines vulnerability assessments, particularly in enterprise environments where security decisions need to be well-justified. As 6G networks introduce AI-driven decision-making, interpretability becomes crucial for ensuring transparency, accountability, and trust in automated cybersecurity solutions.

\subsubsection{Automation and Scalability Bottlenecks}

With the rapid growth of IoT, cloud-native applications, and AI-powered networks, vulnerability detection must scale to secure trillions of interconnected devices. Traditional security tools struggle to automate and scale assessments in ultra-dense 6G environments~\cite{survey}. Signature-based intrusion detection, manual penetration testing, and rule-based anomaly detection require constant updates, human intervention, and extensive configuration, making them inadequate for dynamic cyber threat landscapes. The rise of AI-generated malware, adversarial attacks, and quantum cryptographic threats heightens the need for scalable and self-adaptive security solutions.

The limitations underscore the need for an adaptive, explainable, and reliable approach to vulnerability detection. GAI offers new possibilities by utilizing deep learning, natural language processing, and advanced reasoning to enhance cybersecurity. By improving coverage for novel threats, incorporating contextual reasoning, providing explainable security assessments, and automating large-scale operations, GAI-enabled models represent a transformative shift in cybersecurity strategies. The next section examines how these models can address the constraints of traditional methods and reshape vulnerability detection in the 6G era.
\section{Generative AI-powered Vulnerability Detection}
\label{sec: method}

\begin{figure*}
    \centering
    \includegraphics[width=1\linewidth]{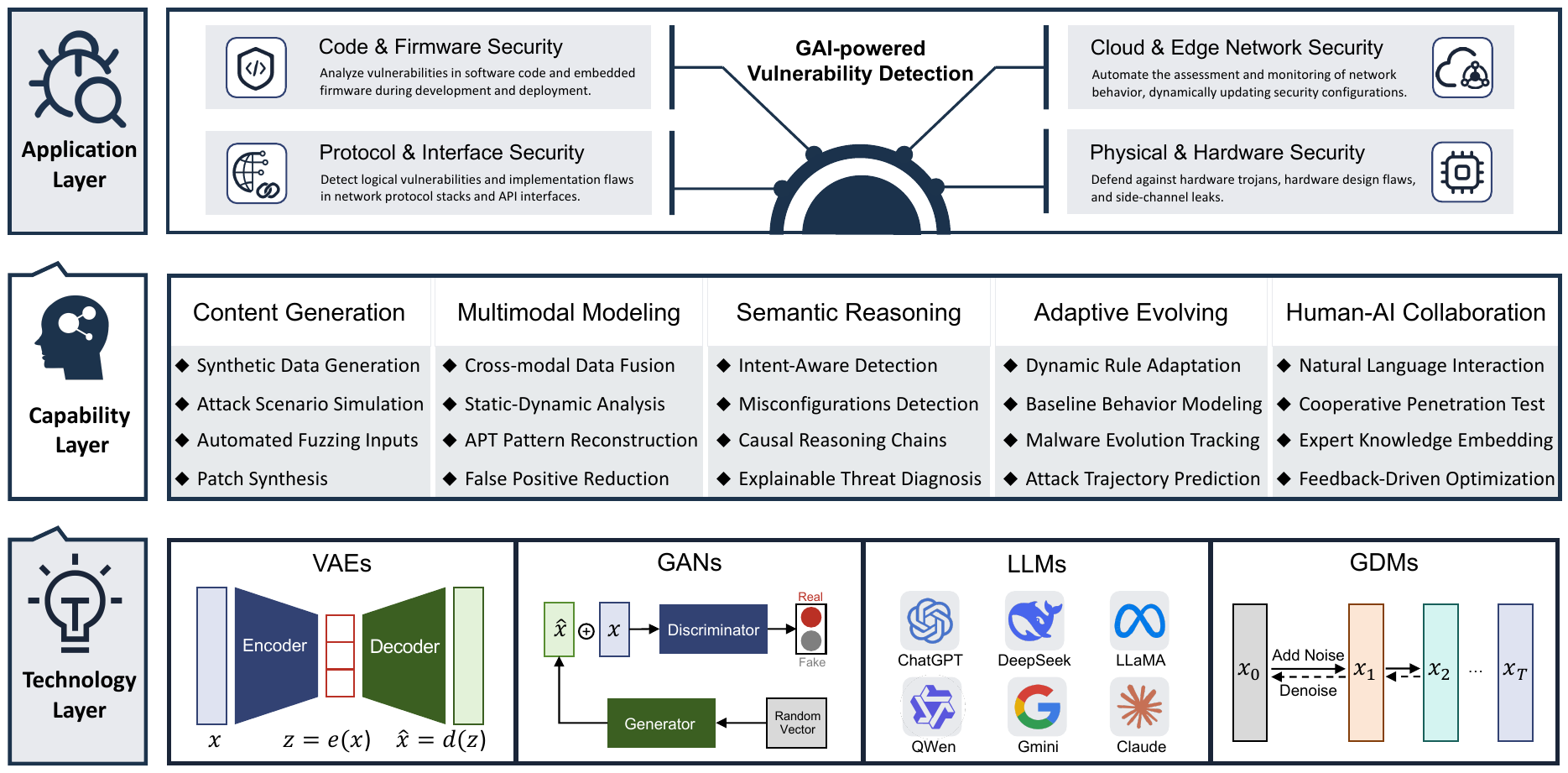}
    \caption{Three-layer architecture of GAI-enabled vulnerability detection}
    \label{fig: gai4vd}
\end{figure*}

With the rapid evolution of wireless networks, traditional vulnerability detection methods struggle against increasingly sophisticated cyber threats. GAI offers a transformative approach by using generative models to generate attack scenarios, analyze evolving threats, and enhance detection efficiency. This section introduces a three-layer architecture: the Technology Layer, the Capability Layer, and the Application Layer, as illustrated in Fig.~\ref{fig: gai4vd}.

\subsection{Technology Layer}

The Technology Layer serves as the foundation of GAI-powered vulnerability detection, integrating advanced generative models to enhance security analysis. These models enable adaptive learning, adversarial simulation, and automated anomaly detection, making them crucial for securing wireless networks and distributed infrastructures.

\subsubsection{Variational Autoencoders (VAEs)}

VAEs encode input data into a latent space representation and reconstruct it to model normal system behavior. By capturing deviations from expected patterns, VAEs effectively detect anomalies in network traffic, system logs, and firmware. Their ability to identify irregularities makes them well-suited for wireless security applications, such as detecting backdoor insertions in firmware and monitoring IoT device communications. Studies have shown that VAEs can prevent firmware-level attacks by analyzing system integrity and flagging abnormal behavioral patterns~\cite{firmware}.

\subsubsection{Generative Adversarial Networks (GANs)}

GANs operate through a generator-discriminator framework, where the generator creates synthetic data and the discriminator distinguishes between real and synthetic samples. This adversarial training allows GANs to simulate attack scenarios, test security defenses, and enhance threat detection. In wireless security applications, GANs generate adversarial inputs to bypass detection systems and improve penetration testing by creating realistic exploit payloads~\cite{edge}. GANs also play a crucial role in stress-testing network protocols for proactive threat mitigation in wireless communications~\cite{survey}.

\subsubsection{Large Language Models (LLMs)}

LLMs, such as GPT-4, Deepseek, and LLaMA, analyze code semantics, configuration files, and network protocol specifications to detect vulnerabilities with contextual awareness. Unlike static analyzers, LLMs understand cross-file dependencies, interpret API misuse, and identify security misconfigurations in wireless networks. For instance, fine-tuned LLMs for code auditing can automatically detect software vulnerabilities and generate patches, enhancing secure software development~\cite{grace}. Additionally, LLMs are vital in cyber threat intelligence, extracting insights from security reports, classifying attack patterns, and predicting emerging threats~\cite{cloud}.

\subsubsection{Generative Diffusion Models (GDMs)}

GDMs iteratively add and remove noise from data, refining its structure over time. These models excel at learning complex patterns and generating high-fidelity representations, making them useful for code vulnerability detection. For instance, GDMs can generate fine-grained code representations from graph-level program slices and predict vulnerable nodes. By modeling code structure and semantic relationships, GDMs enhance vulnerability localization, identifying subtle flaws that traditional static or dynamic analyses may miss~\cite{fvddpm}. This approach significantly improves vulnerability detection in large-scale software projects, wireless protocol stacks, and embedded firmware by accurately reconstructing critical vulnerability points.

In addition to these generative models, reinforcement learning and neural-symbolic AI further enhance GAI’s capabilities, enabling adaptive security automation and real-time vulnerability detection in 6G-enabled wireless networks~\cite{trojanforge}.

\subsection{Capability Layer}

The Capability Layer defines the core functionalities of GAI in vulnerability detection, transforming its technological foundations into practical security solutions. This layer introduces five critical capabilities: Content Generation, Multimodal Modeling, Semantic Reasoning, Adaptive Evolving, and Human-AI Collaboration.

\subsubsection{Content Generation}

Content generation is a defining strength of GAI, allowing it to autonomously create synthetic attack scenarios, fuzzing inputs, and security patches. Unlike traditional methods, which rely on static vulnerability databases, GAI simulates adaptive security insights by modeling potential threats before they occur.

GAI-generated synthetic datasets enhance IDS training, increasing resilience against zero-day attacks. Automated fuzzing techniques expose vulnerabilities in wireless communication protocols, including 5G/6G, IoT, and SDN interfaces. Furthermore, GAI-driven patch synthesis assists in automated vulnerability remediation, minimizing the need for manual security interventions.

\subsubsection{Multimodal Modeling}

Multimodal modeling integrates diverse security data, such as network traffic, log files, source code, and behavior records, to correlate threats across multiple sources. Traditional security tools, which analyze isolated data points, often miss complex attack patterns.

GAI enhances the detection of protocol misconfigurations, firmware vulnerabilities, and runtime anomalies. In Advanced Persistent Threat (APT) reconstruction, multimodal AI combines historical threat intelligence with real-time network telemetry for comprehensive cyberattack reconstructions. Additionally, GAI reduces false positives by integrating multiple data sources, improving detection accuracy in complex wireless environments.

\subsubsection{Semantic Reasoning}

Semantic reasoning enables context-aware vulnerability detection by understanding system intent, relationships, and implications beyond simple pattern matching. Unlike rule-based systems, GAI-powered reasoning interprets security configurations, policy documents, and source code logic to provide deeper insights into network security risks.

A key application of semantic reasoning is misconfiguration detection, where AI identifies insecure network settings, weak access controls, and API misuse in wireless protocols and edge computing environments. By leveraging causal reasoning chains, GAI can trace attack vectors back to their root causes, improving explainable threat diagnosis. This capability ensures that vulnerability assessments are not only automated but also interpretable, aiding security analysts in making faster, more informed decisions.

\subsubsection{Adaptive Evolving}

Traditional security systems struggle with evolving malware, adversarial attacks, and dynamic network threats due to their reliance on predefined detection rules. GAI addresses this challenge through self-learning, real-time adaptation, and proactive security adjustments. For example, LLMs can leverage Retrieval-Augmented Generation (RAG) technology to provide real-time threat intelligence, delivering the latest insights into vulnerability detection.

Dynamic rule adaptation enables security frameworks to autonomously refine detection policies, preventing AI-generated exploits from bypassing traditional defenses. Baseline behavior modeling continuously learns from network traffic, authentication logs, and cloud activity, improving anomaly detection in 6G and IoT-driven infrastructures. Additionally, malware evolution tracking ensures models can recognize new attack variants, while attack trajectory prediction helps forecast potential attack paths before adversaries can exploit them.

\subsubsection{Human-AI Collaboration}

GAI significantly enhances human-AI collaboration by combining AI’s analytical efficiency with human expertise in cybersecurity decision-making. LLM-driven security assistants generate explainable threat intelligence reports, supporting analysts in penetration testing, incident response, and risk mitigation strategies.

Through natural language interaction, AI copilots support cooperative penetration testing, where GAI models generate adaptive attack simulations while human experts refine security strategies. Additionally, feedback-driven optimization allows AI models to continuously improve by incorporating expert security insights, ensuring cybersecurity frameworks remain contextually relevant and adaptive.

\subsection{Application Layer}

The Application Layer illustrates the transformative role of GAI in securing wireless networks, encompassing software, protocols, cloud-edge infrastructures, and hardware components. With its comprehensive generation, understanding, and reasoning ability, GAI enhances vulnerability detection across 6G networks, IoT ecosystems, and cyber-physical systems. The recent representative GAI-powered vulnerability detection methods are summarized in Table.~\ref{tab: gai4vd}.

\begin{table*}[]
\small
\caption{Overview of representative GAI-powered Vulnerability Detection Methods.}
\label{tab: gai4vd}
\centering
\begin{tabular}{ccccccccp{8.5cm}}
\toprule
\multirow{2}{*}{\textbf{App.}} & \multirow{2}{*}{\textbf{Ref.}} & \multirow{2}{*}{\textbf{Tech.}} & \multicolumn{5}{c}{\textbf{Capability}} & \multirow{2}{*}{\textbf{Description}} \\  \cmidrule(lr){4-8}
& & & CG & MM & SR & AE & HAC &\\ \midrule
\multirow{3}{*}[-15pt]{CFS} & \cite{grace} & LLMs & \scriptsize\faCircle & \scriptsize\faCircle & \scriptsize\faCircleO & \scriptsize\faCircleO & \scriptsize\faCircleO & Enhance LLM-based software vulnerability detection using graph structural information and in-context learning.\\
& \cellcolor{gray!20}\cite{fvddpm} & \cellcolor{gray!20}GDM & \cellcolor{gray!20}\scriptsize\faCircle & \cellcolor{gray!20}\scriptsize\faCircleO & \cellcolor{gray!20}\scriptsize\faCircleO & \cellcolor{gray!20}\scriptsize\faCircleO & \cellcolor{gray!20}\scriptsize\faCircleO & \cellcolor{gray!20}Use GDMs to generate fine-grained code representations from graph-level program slices and predict vulnerable nodes. \\
& \cite{firmware}& VAEs& \scriptsize\faCircle & \scriptsize\faCircleO & \scriptsize\faCircleO & \scriptsize\faCircleO & \scriptsize\faCircleO & Extracts features from IoT device memory to train VAEs, effectively preventing backdoor exploitation in firmware.\\ \midrule
\multirow{3}{*}[-15pt]{PIS} & \cellcolor{gray!20}\cite{llmif} & \cellcolor{gray!20}LLMs & \cellcolor{gray!20}\scriptsize\faCircle & \cellcolor{gray!20}\scriptsize\faCircleO & \cellcolor{gray!20}\scriptsize\faCircle & \cellcolor{gray!20}\scriptsize\faCircleO & \cellcolor{gray!20}\scriptsize\faCircleO & \cellcolor{gray!20}Use LLMs to automate protocol specification analysis, effectively identifying 11 vulnerabilities in the Zigbee protocol. \\
& \cite{luataint} & LLMs & \scriptsize\faCircle & \scriptsize\faCircleO & \scriptsize\faCircle & \scriptsize\faCircleO & \scriptsize\faCircleO & Combine static taint analysis with a LLM to automate vulnerability detection for IoT device web configuration interfaces.\\
& \cellcolor{gray!20}\cite{api} & \cellcolor{gray!20}LLMs & \cellcolor{gray!20}\scriptsize\faCircle & \cellcolor{gray!20}\scriptsize\faCircleO & \cellcolor{gray!20}\scriptsize\faCircle & \cellcolor{gray!20}\scriptsize\faCircleO & \cellcolor{gray!20}\scriptsize\faCircleO & \cellcolor{gray!20} Use LLMs to generate API parameter security rules for detecting API misuse from incorrect parameter usage.\\ \midrule
\multirow{3}{*}[-15pt]{CES} & \cite{cloud} & LLMs & \scriptsize\faCircle & \scriptsize\faCircleO & \scriptsize\faCircle & \scriptsize\faCircleO & \scriptsize\faCircle & Employ RAG and CWE as an external knowledge base to enhance LLMs' vulnerability detection and analysis capabilities. \\
& \cellcolor{gray!20}\cite{idsagent} & \cellcolor{gray!20}LLMs & \cellcolor{gray!20}\scriptsize\faCircle & \cellcolor{gray!20}\scriptsize\faCircle & \cellcolor{gray!20}\scriptsize\faCircle & \cellcolor{gray!20}\scriptsize\faCircle & \cellcolor{gray!20}\scriptsize\faCircle & \cellcolor{gray!20}Propose an LLM-powered agent to facilitate IDS pipeline, improving efficiency and detection of zero-day attacks. \\
& \cite{edge} & GANs & \scriptsize\faCircle & \scriptsize\faCircleO & \scriptsize\faCircleO & \scriptsize\faCircleO & \scriptsize\faCircleO & 
Use GANs to generate synthesized samples, thereby expanding the dataset and enhancing performance on minor categories. \\ \midrule
\multirow{3}{*}[-15pt]{PHS}  & \cellcolor{gray!20}\cite{trojanforge}  & \cellcolor{gray!20}GANs & \cellcolor{gray!20}\scriptsize\faCircle & \cellcolor{gray!20}\scriptsize\faCircleO & \cellcolor{gray!20}\scriptsize\faCircleO & \cellcolor{gray!20}\scriptsize\faCircleO & \cellcolor{gray!20}\scriptsize\faCircleO &
\cellcolor{gray!20}Employ RL and GANs to generate adversarial examples of Hardware Trojans that can effectively bypass detection systems.\\
 & \cite{hardwarebug} & LLMs & \scriptsize\faCircle & \scriptsize\faCircleO & \scriptsize\faCircle & \scriptsize\faCircleO & \scriptsize\faCircle &
Use LLMs to automatically repair identified security-relevant bugs in hardware designs by generating replacement Verilog code.\\
 & \cellcolor{gray!20}\cite{emsim} & \cellcolor{gray!20}GANs & \cellcolor{gray!20}\scriptsize\faCircle & \cellcolor{gray!20}\scriptsize\faCircleO & \cellcolor{gray!20}\scriptsize\faCircleO & \cellcolor{gray!20}\scriptsize\faCircleO & \cellcolor{gray!20}\scriptsize\faCircleO & \cellcolor{gray!20}Propose an electromagnetic side-channel leakage evaluation framework that utilizes a GAN to predict electromagnetic emanations.
 \\ \bottomrule

\multicolumn{9}{l}{\footnotesize APP., Ref. and Tech. are abbreviations of Application, Reference and Technology, respectively. }\\

\multicolumn{9}{l}{\footnotesize CFS-Code \& Firmware Security, PIS-Protocol \& Interface Security, CES-Cloud \& Edge Network Security, PHS-Physical \& Hardware Security.}\\

\multicolumn{9}{l}{\footnotesize CG-Content Generation, MM-Multimodal Modeling, SR-Semantic Reasoning, AE-Adaptive Evolving, HAC-Human-AI Collaboration.}\\

\end{tabular}
\end{table*}

\subsubsection{Code \& Firmware Security}

Code and firmware security is vital for preventing remote code execution, supply chain attacks, and firmware backdoors in wireless network infrastructures. As 6G and SDN architectures expand, resilient and automated security mechanisms are essential.

GAI enhances code auditing, software vulnerability detection, and security patch generation through automated analysis. As demonstrated by DeepCode\footnote{DeepCode: https://snyk.io/platform/deepcode-ai/}, which integrates symbolic AI and GAI to detect security flaws and propose automated fixes. Shao~\textit{et al.}~\cite{fvddpm} introduced FVD-DPM, which formalizes vulnerability detection as a diffusion-based graph-structured prediction problem, achieving both precise vulnerability identification and localization. Similarly, GRACE~\cite{grace} enhances LLM-based software vulnerability detection by incorporating graph structure and contextual learning.

In firmware security, VAEs and GANs have been successfully applied to detect anomalous memory modifications, ensuring secure firmware deployment in IoT and edge devices. Iqbal~\textit{et al.}~\cite{firmware} demonstrated that memory-based feature extraction with VAEs effectively prevents firmware backdoor exploitation. These techniques reinforce security of software and firmware in wireless networks.

\subsubsection{Protocol \& Interface Security}

Wireless protocols, like Wi-Fi, Bluetooth, and IoT communication stacks, are vulnerable to attacks such as man-in-the-middle exploits, authentication flaws, and denial-of-service (DoS) incidents. Protecting these interfaces is critical to maintaining network integrity, confidentiality, and availability.

GAI, particularly LLMs, enhances protocol and interface security through automated fuzz testing, API verification, and configuration optimization. Wang~\textit{et al.}~\cite{llmif} introduced LLMIF, which leverages LLMs for automated fuzz testing of IoT protocols, uncovering previously unknown vulnerabilities in Zigbee networks. To address the security challenges posed by diverse IoT web interfaces, Xiang~\textit{et al.}~\cite{luataint} proposed LLM-assisted taint analysis to improve vulnerability detection across IoT and web-based interfaces. Liu~\textit{et al.}~\cite{api} utilized LLMs for API security analysis, detecting improper parameter usage that could lead to vulnerabilities.

By incorporating GAI-driven fuzzing and automated verification, wireless infrastructures achieve proactive and efficient vulnerability assessments, minimizing risks stemming from misconfigurations, insecure API usage, and weak authentication mechanisms.

\subsubsection{Cloud \& Edge Network Security}

Cloud and edge computing serve as the foundation of 6G networks, industrial IoT, and smart city infrastructures but introduce significant attack surfaces, including misconfiguration, intrusions, and privilege escalation. Given the complexity of multi-cloud and edge architectures, intelligent security automation is essential for effective threat mitigation.

IBM’s Autonomous Cloud Security (ASC)\footnote{ASC: https://www.ibm.com/services/autonomous-security-cloud} exemplifies the application of GAI for cloud vulnerability detection, policy enforcement, and dynamic risk modeling. Threat intelligence enhances GAI capabilities through external knowledge integration. LLM-CloudSec~\cite{cloud} incorporates RAG and the Common Weakness Enumeration (CWE) database into LLMs, enabling unsupervised, fine-grained vulnerability classification in cloud security assessments. Additionally, IDS-Agent~\cite{agent} employs LLMs with a knowledge base, security tools, and machine learning models for real-time intrusion detection in edge networks, demonstrating strong performance against zero-day attacks. A study by~\cite{edge} further demonstrated that GAI-driven adversarial learning enhances anomaly detection, improving security resilience in 6G-powered edge infrastructures.

In summary, GAI enables autonomous threat assessment and mitigation, reducing manual intervention while strengthening security across distributed and federated cloud environments.

\subsubsection{Physical \& Hardware Security}

Securing hardware components—including 6G base stations, IoT sensors, and network routers—is fundamental to wireless network resilience. Hardware vulnerabilities, such as hardware trojans, and electromagnetic side-channel attacks and insecure chip designs, pose serious threats to global communications infrastructure.

GAI improves hardware security through automated chip verification, adversarial attack simulations, and vulnerability detection. TrojanForge, developed by Sarihi~\textit{et al.}~\cite{trojanforge}, employs GANs and reinforcement learning to simulate hardware Trojan insertions, aiding in vulnerability identification. Additionally, LLM-driven hardware verification enhances Verilog-based vulnerability detection, automating compliance checks and design error corrections~\cite{hardwarebug}.

For side-channel attack detection, GANs generate synthetic electromagnetic signatures to improve countermeasures against electromagnetic leakage. The EMSIM+ model~\cite{emsim} applies GAN-powered simulations to predict electromagnetic emissions, facilitating hardware security certification. As hardware and software integration deepens in IoT and cyber-physical systems, GAI ensures that security protections extend beyond software defenses, providing a comprehensive security framework for next-generation wireless infrastructures.
\section{Case Study: LLM-driven Code Vulnerability Detection}
\label{sec: case}

In wireless networks, software vulnerabilities pose significant security risks, allowing attackers to intercept communications, execute remote code, or disrupt services. Recent advancements in pretrained models like CodeT5, CodeBERT, UNIXCoder, and LLaMA provide new opportunities for automated vulnerability detection. This section presents a case study evaluating these models' effectiveness in detecting vulnerabilities and comparing their performance to derive insights for future research.

\subsection{Experiment Setups}
\subsubsection{Problem Definition} Vulnerability detection is formulated as a binary classification task, where a function \( f: C \rightarrow \{0, 1\} \) is learned to map a given code snippet \( c \in C \) to a label indicating the presence or absence of a vulnerability. 

\subsubsection{Datasets} Experiments are conducted on two benchmark datasets. Devign\footnote{Devign Dataset: https://github.com/epicosy/devign} is widely used and covers various vulnerability types, such as buffer overflow, SQL injection, and information disclosure. It contains 27,652 vulnerable functions and 31,313 benign functions. PrimeVul\footnote{PrimeVul Dataset: https://github.com/DLVulDet/PrimeVul} is a recently introduced comprehensive vulnerability detection benchmark. PrimeVul is highly imbalanced, with 228,800 benign functions and 6,968 vulnerable functions. It includes numerous wireless network-related software and tools, such as tcpdump, NetworkManager, open5gs, and aircrack-ng.

\subsubsection{Models and Implementation} 

We compare two traditional methods and four open-source LLMs of varying sizes, as detailed in Table~\ref{tab: model}. All experiments are conducted on an NVIDIA H800 GPU with an AMD EPYC 9654 96-Core Processor. Devign, a classical graph-based model, is implemented using Jeron for transforming the code to graphs. Additionally, a Word2Vec tokenizer combined with an MLP classifier serves as a traditional baseline.

\begin{table}[h]
\centering
\caption{Model Comparison.}
\label{tab: model}
\begin{tabular}{c|c|c|c}
\toprule
\textbf{Type}                                & \textbf{Model}       & \textbf{Parameters} & \textbf{Release Year}   \\ \midrule
\multirow{2}{*}{Traditional Method}   & Devign       & 1M         &    2019 \\
                                     & Word2Vec+MLP & 36M          &    -    \\ \midrule
\multirow{3}{*}{\begin{tabular}[c]{@{}c@{}}Domain Specific-\\ Pretrained Model\end{tabular}} & CodeT5       & 60M        & 2021 \\
                                     & CodeBERT     & 125M       & 2020 \\
                                     & UNIXCoder    & 125M       & 2022 \\
                                     \midrule
Genreal LLM                          & LLaMa 3.1-8B & 8B         & 2024 \\ \bottomrule
\end{tabular}
\end{table}

\begin{figure}
    \centering
    \includegraphics[width=1\linewidth]{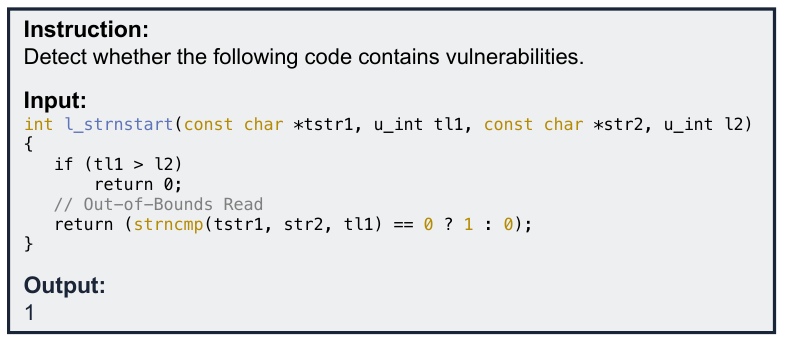}
    \caption{Prompt Template for Large Language Models.}
    \label{fig: prompt}
\end{figure}

For CodeT5\footnote{CodeT5: Identifier-aware Unified Pre-trained Encoder-Decoder Models for Code Understanding and Generation (Wang et al., EMNLP 2021)}, CodeBERT\footnote{CodeBERT: A Pre-Trained Model for Programming and Natural Languages (Feng et al., ACL 2020)}, and UNIXCoder\footnote{UniXcoder: Unified Cross-Modal Pre-training for Code Representation (Guo et al., ACL 2022)}, all fine-tuning experiments follow a standardized framework aligned with existing benchmarks, using a fixed learning rate of \(2 \times 10^{-5}\) over 10 epochs. For LLaMA\footnote{The Llama 3 Herd of Models.(Grattafiori et al., arXiv 2024)}, we configure the model parameters using the default settings provided by Meta AI and employ Low-Rank Adaptation\footnote{LoRA: Low-rank adaptation of large language models. (Hu et al., ICLR 2022)} for instruction tuning. AdamW is used as the optimizer with a learning rate of 1e-4, and the model is trained for 5 epochs. We adapt the template provided by Alpaca\footnote{Alpaca: https://crfm.stanford.edu/2023/03/13/alpaca.html}. As illustrated in Fig.~\ref{fig: prompt}, the vulnerable example is extracted from the tcpdump project\footnote{CVE-2017-13010: https://nvd.nist.gov/vuln/detail/CVE-2017-13010}, where an improper function call leads to an out-of-bounds memory read, potentially causing crashes or exposing sensitive information.

\subsubsection{Evaluation Metrics} To ensure a fair assessment under imbalanced data conditions, Accuracy (ACC) and F1-score (F1) are reported as key evaluation metrics. Additionally, inference efficiency is measured by comparing the number of samples processed per second, providing practical insights for real-world deployment.

\subsection{Main Results}

\begin{table}[h]
\caption{Performance and efficiency of models on Benchmark Datasets.}
\label{tab: case}
\centering
\begin{tabular}{c|c|c|c|c}
\toprule
\textbf{Dataset }                  & \textbf{Model}        & \textbf{ACC}   & \textbf{F1}    & \textbf{Samples/s} \\ \midrule
\multirow{6}{*}{Devign}   & Devign       & 51.01 & 45.54  &   1,294        \\
                          & Word2Vec+MLP & 54.47 & 51.82 &  67,424\\ 
                          & CodeT5       & 62.63 & 58.10  &  1153   \\
                          & CodeBERT     & 64.39 & 65.58 &  351  \\
                          & UNIXcoder    & 61.09 & \textbf{65.98} & 344   \\        \\ 
                          & LLaMa 3.1-8B & \textbf{65.05} & 55.08 &   15        \\ \midrule
\multirow{6}{*}{Primevul} & Devign       & 98.05 & 4.35  &   1,613        \\
                          & Word2Vec+MLP & 97.64 & 15.85 &  144,833  \\
                          & CodeT5       & 97.67 & \textbf{18.60}  &  1,740   \\
                          & CodeBERT     & 97.37 & 17.45 &  355  \\
                          & UNIXcoder    & 97.29 & 18.07 & 317 \\
                          & LLaMa 3.1-8B & \textbf{98.44} & 13.33 &   16       \\ \bottomrule
\end{tabular}
\end{table}

Table~\ref{tab: case} summarizes the model performances. Fine-tuned LLMs leverage pre-learned representations, achieving superior F1 compared to traditional deep learning methods, albeit with increased inference latency. Notably, general LLMs such as LLaMA do not outperform domain-specific pretrained models like CodeBERT, even when LLaMA has more parameters. UNIXCoder achieves the highest performance on Devign, benefiting from pre-training objectives that enhance its understanding of code logic and control flow. However, due to the severe class imbalance in PrimeVul, the F1 across all models are significantly lower than on Devign. These results highlight the potential of LLMs in code vulnerability detection while emphasizing the need to balance model size, dataset characteristics, and computational efficiency.

\subsection{Discussion}
The findings from our case study underscore several critical considerations in applying LLMs to vulnerability detection in wireless networks.

\subsubsection{Integration of Domain Knowledge Enhances Detection}

Domain-specific LLMs outperform general models, underscoring the importance of incorporating code-specific knowledge. Pre-training alone lacks sufficient security insights, whereas integrating structured security data improves a model’s ability to recognize semantic patterns of vulnerabilities, reducing false positives and enhancing detection reliability.

\subsubsection{Data Imbalance Compromises Model Performance} 

Real-world software projects contain disproportionately fewer vulnerable functions, resulting in highly imbalanced training data. Models trained on such datasets tend to be biased toward benign samples, leading to poor recall in detecting vulnerabilities. Addressing this issue requires techniques such as data augmentation, re-weighted loss functions, or oversampling methods to ensure robust vulnerability detection.

\subsubsection{Trade-off between performance and efficiency}

Although fine-tuned LLMs achieve superior detection performance, their computational demands pose challenges for real-time deployment in resource-constrained environments. This trade-off suggests that future research should explore lightweight model optimization techniques, such as adaptive quantization and knowledge distillation, to enable efficient vulnerability detection in wireless networks.
\section{Future Research Directions}
\label{sec: future}

While GAI has shown significant promise in enhancing vulnerability detection for 6G wireless networks, its potential remains largely untapped. Several research challenges must be addressed to ensure its effectiveness, security, and scalability. Future efforts should focus on improving model efficiency, content authenticity, adversarial robustness, knowledge integration, and privacy-preserving mechanisms.

\subsection{Lightweight Generative Models}

Deploying GAI-powered vulnerability detection in 6G wireless networks, particularly in edge environments, necessitates the development of lightweight and resource-efficient models. Large-scale transformers and diffusion models are computationally expensive, making them impractical for real-time security applications in constrained environments. Future research should explore model compression techniques such as knowledge distillation, pruning, and quantization to reduce computational overhead while maintaining detection accuracy. Additionally, the design of edge-native architectures optimized for distributed security tasks can improve inference speed and reduce latency, enabling real-time vulnerability detection in ultra-dense 6G networks.

\subsection{High-Authenticity Content Generation}

The scarcity of benchmark datasets for wireless network security, particularly real vulnerability data, hinders the advancement of vulnerability detection. GAI can address this limitation by synthesizing attack scenarios, adversarial traffic, and vulnerable samples, thereby improving vulnerability models. However, ensuring the authenticity of generated data remains a challenge, as synthetic samples must accurately reflect real-world threats. Future research should refine verification techniques, such as adversarial training and mixed data augmentation, to enhance data reliability. Additionally, LLMs can generate interpretable reports and strategies, however, they often produce misleading or unrealistic content due to hallucinations. Therefore, addressing this issue is crucial for maintaining the accuracy of detection and interpretation.

\subsection{Adversarial Robustness and Model Security}

As GAI becomes increasingly integrated into cybersecurity defenses, its susceptibility to adversarial attacks must be addressed. Attackers can manipulate inputs to deceive GAI models, evade detection, or generate misleading security insights. Future research should explore adversarial training techniques, robust feature extraction mechanisms, and uncertainty-aware models to strengthen resilience against adversarial perturbations. Furthermore, improving the security of model parameters and preventing model inversion attacks will be critical in protecting GAI-driven security frameworks from exploitation. 

\subsection{Integration of External Knowledge and Threat Intelligence}

While GAI models learn from extensive datasets, they often struggle to dynamically incorporate evolving cybersecurity intelligence. Future research should focus on integrating real-time threat intelligence and domain knowledge into generative models using techniques such as RAG, knowledge graph embedding, and continual learning. This would enable GAI systems to continuously update their security insights based on emerging attack trends and vulnerabilities. Moreover, combining structured threat databases (e.g., MITRE ATT\&CK, CVE) with LLM-based security reasoning can enhance contextual awareness in vulnerability detection.

\subsection{Privacy-Preserving Generative AI}

Incorporating GAI into wireless network security requires careful consideration of privacy risks. Generative models trained on sensitive security data could inadvertently reveal confidential information or generate biased predictions. Future research should prioritize privacy-preserving techniques, such as differential privacy, federated learning, and secure multiparty computation, to ensure data confidentiality. Additionally, developing decentralized generative models that can operate within privacy-compliant frameworks will be essential for securing 6G networks while maintaining user trust. Ensuring that GAI-generated security reports and vulnerability assessments do not expose sensitive information will also be a key challenge for regulatory compliance.
\section{Conclusion}
\label{sec: conclusion}
GAI presents a transformative solution to the evolving security challenges in 6G wireless networks, alleviating the limitations of traditional vulnerability detection through data synthesis, multimodal reasoning, and adaptive learning. This paper introduced a three-layer framework to systematically explore GAI’s role in enhancing security across software, protocols, cloud-edge infrastructures, and hardware. A case study on LLM-driven code vulnerability detection demonstrated the effectiveness of GAI in identifying vulnerabilities. However, challenges remain, including computational costs, adversarial robustness, data scarcity, and the need for trustworthy GAI-generated content. Future research should focus on lightweight models, high-authenticity data generation, external knowledge integration, and privacy-preserving technologies to enable adaptive, explainable, and reliable cybersecurity solutions. We hope this paper inspires more researchers in the network community to explore the deployment of GAI in securing next-generation 6G networks against emerging threats.

\bibliographystyle{IEEEtran}
\bibliography{reference}

\vfill

\end{document}